\documentclass[aps,aas_macros,prl,twocolumn,showpacs,superscriptaddress,groupedaddress,nofootinbib]{revtex4}  
\usepackage{graphicx}  
\usepackage{dcolumn}   
\usepackage{bm}        
\usepackage{amsmath,amssymb}   
\usepackage{aas_macros}
\usepackage{multirow}
\usepackage{color}
\usepackage{times}
\usepackage{verbatim}
\usepackage{hyperref}
\newcommand{\mr}{\mathrm}
\newcommand{\Mpc}{\ensuremath{\,{\rm Mpc}}}
\newcommand{\K}{\ensuremath{\, {\rm K}}}

\newcommand{\eV}{\ensuremath{\,{\rm eV}}}

\newcommand{\nur}{\ensuremath{{\nu_R}}}
\newcommand{\nul}{\ensuremath{{\nu_L}}}
\newcommand{\los}{l.o.s.}

\begin{document}
\widetext

\title{Probing Neutrino Hierarchy and Chirality via Wakes}

\author{Hong-Ming Zhu} 
\affiliation{Key Laboratory for Computational Astrophysics,  National Astronomical Observatories, Chinese Academy of Sciences, 20A Datun Road, Beijing 100012, China}
\affiliation{University of Chinese Academy of Sciences, Beijing 100049, China}

\author{Ue-Li Pen} 
\affiliation{Canadian Institute for Theoretical Astrophysics, 60 St. George Street, Toronto, Ontario M5S 3H8, Canada}
\affiliation{Canadian Institute for Advanced Research, CIFAR Program in Gravitation and Cosmology, 
Toronto, Ontario M5G 1Z8, Canada}

\author{Xuelei Chen}
\affiliation{Key Laboratory for Computational Astrophysics,  National Astronomical Observatories, Chinese Academy of Sciences, 20A Datun Road, Beijing 100012, China}
\affiliation{University of Chinese Academy of Sciences, Beijing 100049, China}
\affiliation{Center of High Energy Physics, Peking University, Beijing 100871, China}

\author{Derek Inman} 
\affiliation{Canadian Institute for Theoretical Astrophysics, 60 St. George Street, Toronto, Ontario M5S 3H8, Canada}
\date{\today}

\begin{abstract}
The relic neutrinos are expected to acquire a bulk relative velocity with 
respect to the dark matter at low redshifts, and neutrino wakes are expected to develop 
downstream of the dark matter halos.  We propose a method of 
measuring the neutrino mass based on this mechanism.  
This neutrino wake will cause a dipole distortion of the galaxy-galaxy
lensing pattern.  This effect could be detected by combining upcoming lensing 
surveys with a low redshift galaxy survey or 
a 21 cm intensity mapping survey, which can map the neutrino flow 
field.  The data obtained with LSST and Euclid 
should enable us to make a positive detection if the three
neutrino masses are quasidegenerate with each neutrino mass of $\sim$0.1 eV, 
and a future high precision 21 cm lensing survey would allow the normal hierarchy and inverted 
hierarchy cases to be distinguished, 
and even the right-handed Dirac neutrinos may be detectable.
\end{abstract}

\maketitle

{\it Introduction.}---The squared mass differences of the three neutrino species 
have been measured from neutrino oscillation experiments, but 
the individual masses are still unknown. Based on the measured mass square 
differences, the neutrino masses may form the so called normal 
hierarchy ($m_1 \sim m_2 \ll m_3$) or the inverted 
hierarchy ($m_3 \ll m_1 \sim m_2$), or  they may be quasidegenerate
($m_1 \sim m_2 \sim m_3$) \cite{2014ChPhC..38i0001O}. 
Determining the neutrino mass hierarchy is a very
important problem in modern physics \cite{2013arXiv1307.5487C}. 
The suppression of the matter power spectrum on small scales  
provides a way to measure the sum of neutrino 
masses \cite{Bond:1980,Hu:1997}, complementary 
to particle physics experiments, and in some cases 
the combination of these two approaches may
allow the determination of the mass hierarchy \cite{Dodelson:2014}. 

Recently, it was shown that  a bulk relative velocity field between 
neutrinos and cold dark matter (CDM)
exists, with coherent flows over several Mpc \cite{Zhu:2013}.
This produces a cross-correlation dipole between CDM and neutrinos on large
scales. While on nonlinear scales,
as neutrinos flow over dark matter halos, they are gravitationally focused
into a wake.  This downstream excess can be observed
through gravitational lensing. This wake is unique to
neutrinos as the CDM-baryon relative velocity is far too small to
mimic the effect.  The full three-dimensional relative velocity field can be
computed from the galaxy density field using linear perturbation 
theory with good accuracy.
By exploiting the asymmetry of the relative velocity, the neutrino wakes
can be isolated and used to determine the neutrino masses.

Besides the neutrino masses, the very nature of the neutrinos, i.e., whether
they are Majorana or Dirac particles, is still unknown.
If neutrinos are Majorana particles,
the lepton number conservation law is broken, and this may induce neutrinoless
$\beta$ decay, but so far such decays have not been detected despite many
search efforts \cite{2013arXiv1310.4340D}. 
This question may be answered if the big bang relic neutrinos can be detected
with a tritium capture experiment \cite{2013arXiv1307.4738B}, for
the total capture rate in the Majorana case could be twice 
as large as that of the Dirac case \cite{2014JCAP...08..038L}. 
However, detecting the relic neutrinos is very difficult due to their small kinetic energy, 
and furthermore, even if the relic neutrinos are detected, this effect is confounded by the fact that
the local neutrino overdensity due to gravitational clustering is unknown. 
Here, we note that for Dirac neutrinos there are 
right-handed neutrinos ($\nu_R$'s) distinct from the left-handed ones ($\nu_L$'s), and
they could have been produced during the big bang. In many beyond the standard model 
theories, e.g., the models with a heavier right-handed coupling gauge boson $Z^\prime$ \cite{2009RvMP...81.1199L}, 
the left-right symmetry is restored at high energy, then in the very early Universe there would be 
a thermalized $\nu_R$ background in addition to the $\nu_L$ one. However, the left-right 
symmetry must be broken at low energy, and as the $Z^\prime$  is heavier than the $Z$ boson,
the primordial $\nu_R$'s  would decouple before the $\nu_L$'s, and evolve as a
separate relic background. At a later time, these neutrinos
would propagate in mass eigenstates, with both right
and left (regenerated from Yukawa coupling) components, 
but their temperature and number density would be 
distinct from the primordial $\nu_L$ background, which decoupled much later. 
Since the relative velocity with respect to CDM depends on the initial velocity
dispersion \cite{Zhu:2013},  
the relative velocity field of the $\nu_R$ including its magnitude and 
direction would also be different, thus enabling a new
way to answer this important question. (In the seesaw models 
of neutrino masses \cite{2008IJMPA..23.4255X}, 
the light $\nu_L$'s are primarily Majorana particles. 
There are also right-handed neutrinos in such models, but they 
are very heavy and should have decayed in the early Universe.)
In this Letter we discuss how these neutrino wakes are produced by
nonlinear CDM halos and compute the expected lensing signals, 
and investigate the observability with upcoming surveys. 
Moreover, we also delineate the evolution of the $\nu_R$'s for the 
case of Dirac neutrinos, and consider the detectability with this method.

\begin{figure}[tbp]
  \begin{center}
    \includegraphics[width=0.48\textwidth]{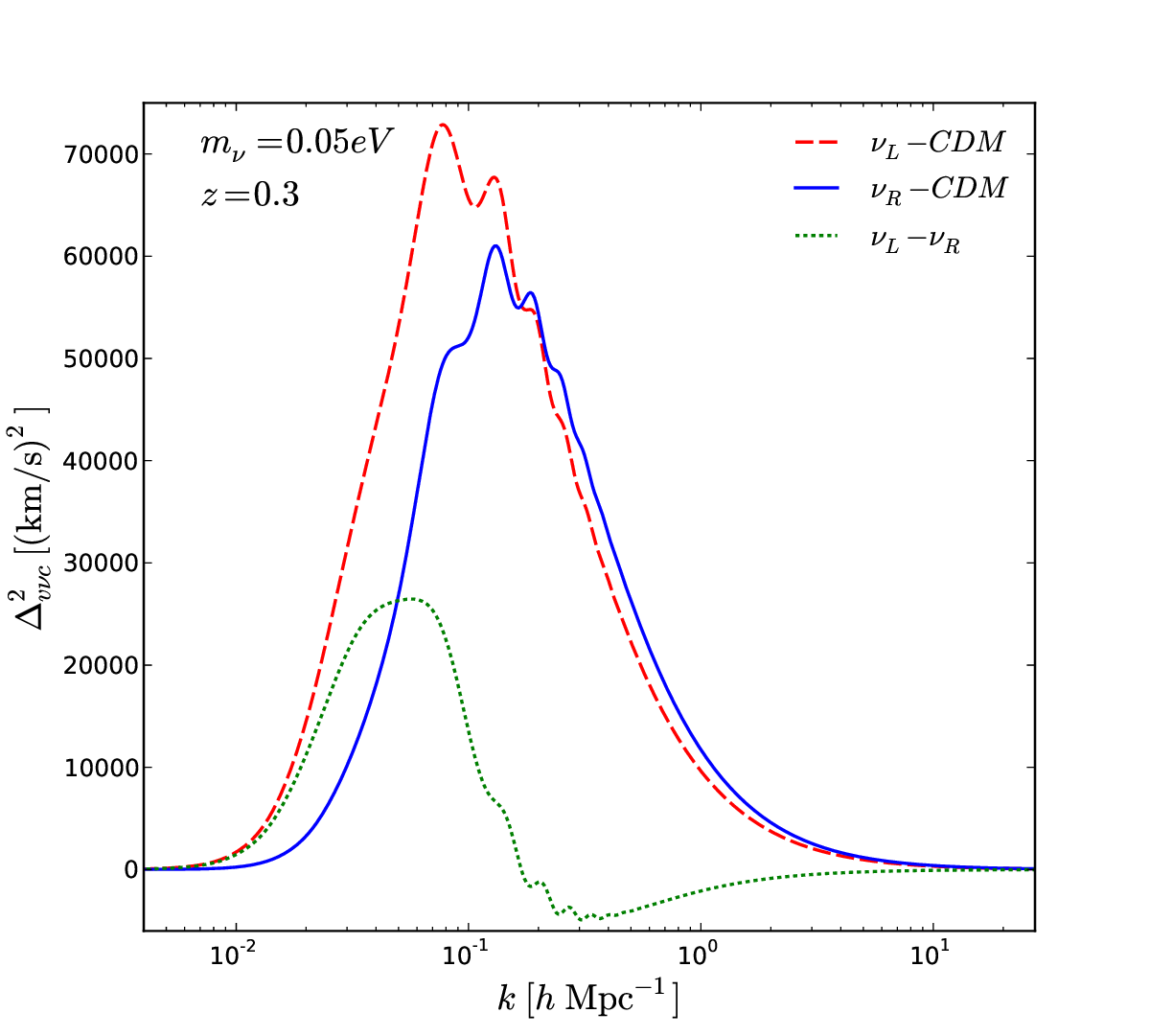}
  \end{center}
  \vspace{-0.7cm}
  \caption{Relative velocity power spectra between the $\nu_L$, $\nu_R$, and
    CDM at $z=0.3$.}
  \label{fig:relvel}
\end{figure}

{\it Neutrino wakes and lensing signal.}---Because of the free streaming of 
neutrinos over large scales, the bulk flow of neutrinos grows slowly compared
to that of CDM;
as a result a bulk relative velocity field between neutrinos and 
CDM is induced.
The relative velocity field can be computed from the primordial density
perturbations and its variance is given by 
\begin{eqnarray}
\label{v_nuc}
\langle v^2_{\nu c}\rangle(z)
&=& \int \frac{dk}{k}\Delta^2_{\zeta}(k)\bigg[ \frac{\theta_\nu (k, z) - \theta_c (k, z)}{k} \bigg]^2, 
\end{eqnarray}
where $\theta\equiv\nabla\cdot{\bm v}$ is the velocity divergence
transfer function and $\Delta^2_\zeta(k)$ is the primordial curvature
power spectrum \cite{Zhu:2013}.
This variance was computed for the $\nu_L$ in Ref. \cite{Zhu:2013} where it was
found to be comparable with the thermal velocity at late times. The $\nu_L$
relative velocity power spectrum is shown in Fig. \ref{fig:relvel} for the
case of a single 0.05 eV neutrino.
The velocity coherent scale, which is defined as the distance where 
the relative velocity correlation function drops to half its maximum value 
\cite{Zhu:2013},
is $R=14.5$ Mpc/$h$ for the $\nu_L$ at $z=0.3$. 
In the following calculations, 
we take $\sqrt{\langle v^2_{\nu_L c}\rangle}=418\ \mr{km/s}$ at $z=0.3$ 
as the typical relative velocity. The one-dimensional neutrino velocity 
dispersion is $\sigma(z)=2.077T_{\nu}(z)/m_\nu\simeq2716\ \mr{km/s}$ 
for the $\nu_L$ at $z=0.3$.

We now compute the neutrino wake induced by a dark matter halo of
mass $M$ via linear response theory. 
We approximate the neutrinos' distribution as Maxwell Boltzmann, 
which is sufficiently accurate for the relevant densities and temperatures.
Here, we take the halo as the origin point.
Assuming neutrinos
flow over the dark matter halo coherently with a velocity $v_{\nu c}(z)
=\sqrt{\langle v^2_{\nu_L c}\rangle}(z)$, the neutrino density contrast at point
$\bm{r}$ is given by \cite{Binney:2008}
\begin{eqnarray}
  \label{delta_nu}
  \delta_\nu(\bm{r})
  &=&\frac{r_B}{r}
  \mr{exp}\bigg(-\frac{v_{\nu c}^2\mr{sin}^2\vartheta}{2\sigma^2}\bigg)
  \bigg[1+\mr{erf}\bigg(\frac{v_{\nu
      c}\mr{cos}\vartheta}{\sqrt{2}\sigma}\bigg)\bigg] \nonumber \\
  &\approx&\frac{r_B}{r}
  \bigg(1+\frac{2}{\sqrt{\pi}}X\mr{cos}\vartheta-X^2\mr{sin}^2\vartheta\bigg),
\end{eqnarray}
where $r_B(z) = GM/\sigma^2(z)$ is the Bondi radius, $\vartheta$ is the
angle between the relative velocity and $\bm{r}$, and $X(z)=v_{\nu
  c}(z)/[\sqrt{2}\sigma(z)]$. The approximation in 
the second line is better than 95\% for the cases we consider.

\begin{figure}[tbp]
  \begin{center}
    \includegraphics[width=0.44\textwidth]{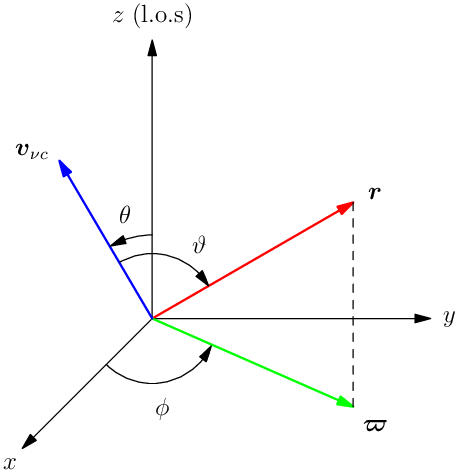}
  \end{center}
  \vspace{-0.7cm}
  \caption{ The coordinate system employed in calculating the lensing
    signal.  $\bm v_{\nu c}$ is the relative velocity, $\bm r$ is the
    point at which we compute the density contrast, and $(\varpi,
    \phi)$ are polar coordinates in the lensing plane.}
  \label{fig:coords}
\end{figure}

With Eq. (\ref{delta_nu}) in hand we can 
compute the expected lensing signal from this wake. The geometry 
is shown in Fig. \ref{fig:coords}.  Taking the $z$ axis
as the direction of our line of sight (\los) and the relative velocity
$\bm{v}_{\nu c}$ to lie in the $x-z$ plane at an angle $\theta$ from
the $z$ axis, then the induced density contrast at any point 
$\bm{r} \equiv (x,y,z)$
can be determined from Eq. (\ref{delta_nu}) via coordinate
transformations.  Here, we also define the polar coordinates $(\varpi, \phi)$ on
the lens ($x-y$) plane, $x=\varpi \cos\phi,\ y=\varpi \sin\phi$, for
later use.  

\begin{figure}[tbp]
\begin{center}
\includegraphics[width=0.48\textwidth]{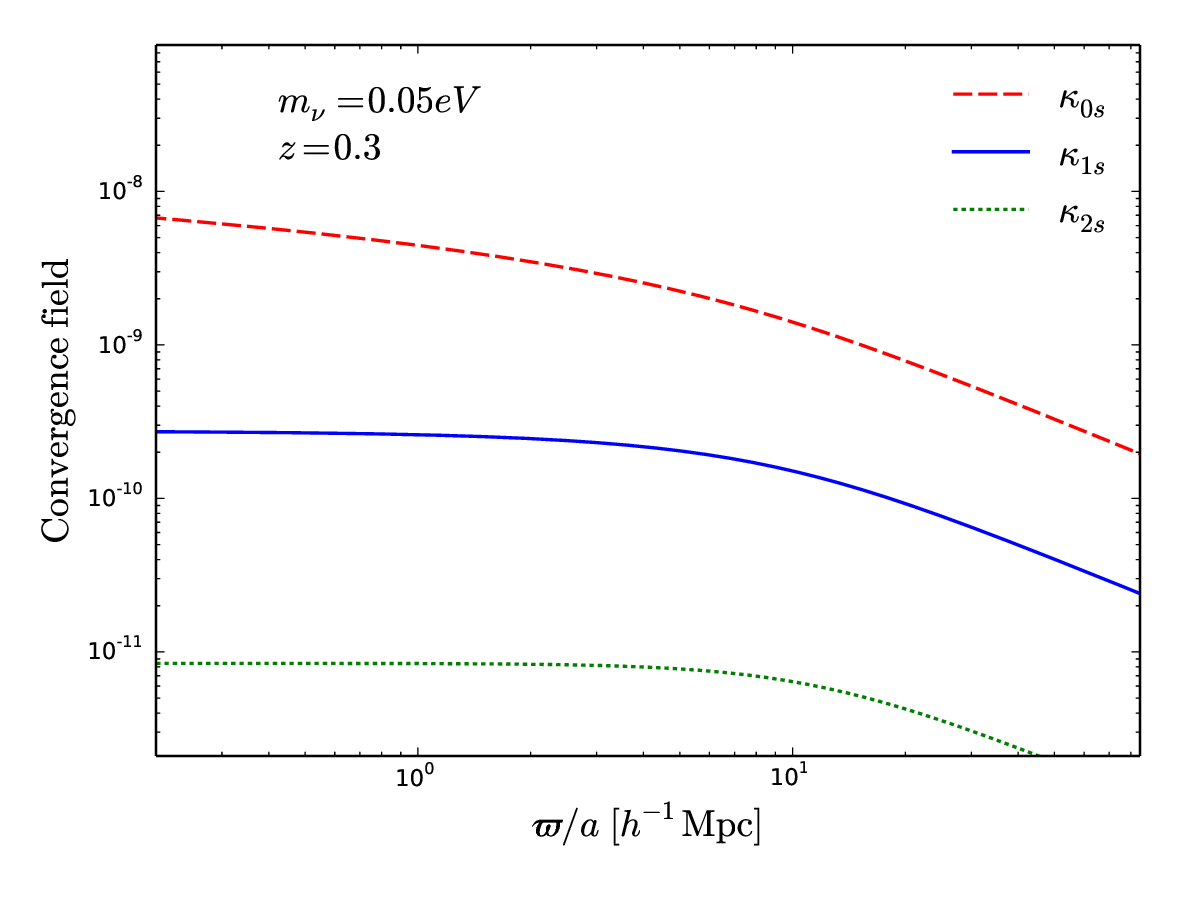}
\end{center}
\vspace{-0.7cm}
\caption{\label{fig:kappa} Different components of the convergence field for 
a single $10^{13} M_\odot$ halo at $z=0.3$, with $m_\nu=0.05$ eV when 
$\theta=\pi/2$. $a=1/(1+z)$ is the scale factor and $\varpi/a$ shows the
comoving distance at redshift $z=0.3$.}
\end{figure}

The perturbed surface neutrino mass density is obtained by integration along the
l.o.s.: $\Sigma_\nu(x,y)=\rho_0\int^{aL/2}_{-aL/2} dz~\delta_\nu(x,y,z)$,
where $\rho_0$ is the unperturbed neutrino mass density, $a$ is the scale 
factor, and $L$ is the
effective coherent scale of the neutrino-CDM relative flow field,
defined as $(4/3)\pi R^3=L^3$.  We find $L$ = 23.4 Mpc/$h$ for
the $\nu_L$ at $z=0.3$ for $m_\nu=0.05\ \mr{eV}$. 
We first consider the contribution from a single halo at $z=0.3$.
Using Eq. (\ref{delta_nu}), we obtain
\begin{eqnarray}
&&  \frac{\Sigma_\nu}{\rho_0r_B}=(1-X^2\sin^2\theta)
  \ln\bigg[\frac{1+\sin\eta}{1-\sin\eta}\bigg]+\frac{4\eta X}{\sqrt{\pi}}\sin\theta\cos\phi
   \nonumber \\
  && +X^2(3\sin^2\theta-2)\sin\eta
  -X^2(\cos^2\theta-1)\sin\eta\cos2\phi,\nonumber
\end{eqnarray}
where $\eta = \arctan(aL/2 \varpi)$.  
The contribution of neutrinos in a redshift slice to the weak lensing 
convergence is given by
$\kappa(\varpi,\phi)=\Sigma_\nu(\varpi,\phi)\omega(\chi)$, where
$\omega(\chi)={4\pi\mr{G}}a\chi\int_z^{\infty}
dz_sn(z_s)(1-\chi/\chi_s)$
is the lensing weight, $\chi$ and $\chi_s$ are comoving distances to the halo
and the source, and $n(z_s)$ is the redshift distribution of the sources 
normalized to unity.
We adopt the source distribution characterized by 
$n(z_s)\propto z_s^\alpha\mr{exp}[-(z_s/z^*)^\beta]$ with $\alpha=2$, $z^*=0.5$,
and $\beta=1$, which corresponds to the Large Synoptic Survey Telescope (LSST) 
survey \cite{Abell:2009}.
The lensing weight peaks at $z=0.38$.
The convergence can be arranged into several multipole terms:
$$\kappa(\varpi,\phi)=\kappa_0(\varpi)+ \kappa_1
(\varpi)\cos\phi+\kappa_2(\varpi)\cos2\phi.$$ 
In Fig. \ref{fig:kappa} we plot the contribution to the different components
of the convergence field from a single halo $\kappa_{0s}$, $\kappa_{1s}$, and 
$\kappa_{2s}$ at $z=0.3$
for $m_\nu=0.05$ eV in the case that the relative velocity is
perpendicular to the l.o.s., i.e., $\theta=\pi/2$.
Here, we take $M=10^{13}M_\odot$, which is the typical mass of a halo.

Although the effect is fairly small, no other effect is known to produce the dipole 
term $\kappa_{1s}$  except for the neutrino wake lensing.
The halo density field itself may have a dipole; however, its contribution 
can be removed by correlating with the relative velocity field 
direction $\langle\delta_h(\bm x)\delta_h(\bm x + \bm r) \hat{\bm r}
\cdot\hat{\bm{v}}_{\nu c}\rangle$,  as it is antisymmetric.  We have
  verified this is the case by $N$-body simulations; without the neutrinos
the measured lensing signal is indeed consistent with zero within 
the numerical accuracy.
The clustering of neutrinos around a growing dark matter halo which moves with
a bulk relative velocity with respect to neutrinos can be calculated similarily 
as in Ref. \cite{2014PhRvD..89f3502L}.

{\it Principle of observation.}---We can reconstruct the relative 
velocity field from the large scale structure 
survey data  $\delta_g$; in Fourier space the velocity is given by
\begin{eqnarray}
\bm{v}_{\nu c} (\bm{k})= \frac{i\hat{\bm{k}}}{k}\frac{\theta_c(k) -
\theta_\nu(k)}{T_c(k)}\frac{\delta_g(\bm{k})}{b_g},
\label{eqn:flinvel}
\end{eqnarray}
where $\theta_c$ and $\theta_\nu$ are the velocity divergence 
transfer functions, $T_c$ is
the density transfer function, and $b_g$ is a linear galaxy bias as 
the galaxy density field is different from that of dark matter. 
The influence of different values of the bias is marginal as we only need
the direction of the relative velocity, which can be reconstructed accurately
\cite{2015PhRvD..92b3502I}.
The lensing dipole can then be calculated with respect to the direction of the 
velocity, so that 
the neutrino wakes add coherently. However, as the neutrino mass is unknown,
we would have to generate the relative velocity fields for many
neutrino masses and determine the neutrino mass as the best fit for
$\kappa_1$. 

In a velocity coherent volume $V=L^3$, each halo produces a wake in
the same direction and adds constructively to the dipole term
($\cos\phi$ has the same value).  For a simple estimate, the number of
halos of typical mass in such a volume is $\bar{n} V$, where
$\bar{n}=\rho_m/M$ and $\rho_m$ is the matter density at the
corresponding redshift. 
Thus, we can multiply the response density by $\bar{n} V$, or
in more sophisticated treatment calculate the integrated response using the 
Press-Schechter model.

There are several coherent velocity volumes along the l.o.s., and 
each of these contributes to the dipole term, but the dipole directions are
randomly oriented with
respect to one another. If the contribution to $\kappa_1$ from each of
these coherent volumes can be measured individually, one may construct
a stacked $\kappa_1$, which will total as the number of coherent volumes
as discussed previously.  
However, it is too difficult to measure $\kappa_1$
for individual velocity coherent volumes. Instead, the total
$\kappa_1$ along each l.o.s. can be observed and the dipoles from
each velocity coherent volume will now add incoherently, resulting that 
$\kappa_1$ grows as $\kappa_1\approx\kappa_{1s}\bar{n}V\sqrt{\chi_\mr{eff}/L}$,
where
$\kappa_{1s}$ was the single halo contribution shown in Fig. \ref{fig:kappa}. For our simple estimate,
we define $\chi_\mr{eff}$ as the comoving length along our l.o.s. where the lensing
weight $\omega(\chi)$ is larger than half of its maximum
value. For the fiducial case here, $\chi_\mr{eff}=2093.4\ \mr{Mpc}/h$. 
Considering the redshift distribution of the lensing halos makes  $\kappa_{1}$ 
about 15\% smaller than this simple estimate.

{\it Right-handed neutrinos.}---If the neutrinos are Dirac type, the 
primordial right-handed neutrinos have the same mass as the left-handed ones, 
but with a different temperature, so that their distribution is also different. The decoupling of 
the $\nu_R$'s depends on the model of their interactions, and 
one has to consider the rich phenomenology and experimental constraints on 
such models \cite{2009RvMP...81.1199L,2014ChPhC..38i0001O}. Here, as a simple
example, we assume that the $\nu_R$'s were 
fully thermalized during the early Universe but decoupled before any
other standard model (SM) particle. This is a plausible case, for the $\nu_R$'s do not participate in 
the SM interactions, and their earlier thermalization must be due to a beyond 
the standard model interaction that freezes out at energies above 1 TeV.
The temperature of these neutrinos can be computed
by ensuring the conservation of entropy density $g_{*S}^\mr{today}T_\gamma^3
= g_{*S}^{\nu_R}T_\nur^3$, where $g_{*S}^{\nu_R}$ is the effective degree of freedom
when the $\nu_R$ decoupled. If the standard model is applicable at such energy
scales except for the addition of right-handed neutrinos, we have 
 $g_{*S}^{\mr{SM}} = 106.75$, and $g_{*S}^{\nu_R}=g_{*S}^{\mr{SM}}$
$+\frac{7}{8}$ (fermions) $ \times 3$ (number of
species) $\times 2$ (particle and antiparticle).  Currently, $g_{*S}
= 3.91$ but again a factor of
$\frac{7}{8}\times6\times(T_{\nu_R}/T_\gamma)^3 $
needs to be added.  This leaves
\begin{equation}
\left(\frac{T_{\nu_R}}{T_\gamma}\right)^3 = \frac{106.75 + \frac{7}{8}\times6}{3.91 +
\frac{7}{8}\times6\times\left(\frac{T_{\nu_R}}{T_\gamma}\right)^3}
\end{equation}
so the extrapolated right-handed neutrino temperature is 
$T_{\nu_R} 
=\left(43/427\right)^{1/3}T_{\nu_L} \simeq 0.905 \K .$
With $\nu_R$'s at this temperature, the number of relativistic degrees of
freedom during the nucleosynthesis era 
is $N_\mr{eff} = 3.04 \times [1 + (T_\nur/T_\nul)^4]
\simeq 3.18$.  This is consistent with the current 
big bang nucleosynthesis constraints \cite{2013PhLB..718.1162A}.
Note that the $\nu_R$ density might be greater if its production 
mechanism is nonthermal, but the big bang nucleosynthesis constrains its 
density to be not much higher than the thermal value.
To model $\nu_R$'s, we use the {\tt CLASS} code \cite{ClassCode}
with neutrinos of the
same mass but a lowered temperature.  The results presented here
only depend on the $\nu_R$'s being thermalized and their current
temperature.

The computation of the relative velocity field of the $\nu_R$ 
is similar to the $\nu_L$. However, $\nu_R$'s decoupled earlier with a
lower temperature, i.e., velocity dispersion, so the relative 
velocity is different from that of the $\nu_L$.
The $\nu_L$  and $\nu_R$  relative velocity spectra are shown in
Fig. \ref{fig:relvel}. The coherent scale 
is $R = 10.0$ \Mpc/$h$ for the $\nu_R$ at $z=0.3$. In this model we take
$\sqrt{\langle v^2_{\nu_R c}\rangle} = 373$ km/s as the typical relative
velocity.  The one-dimensional neutrino velocity dispersion is 
$\sigma(z)=2.077T_\nu(z)/m_\nu \simeq 1263\ \mr{km/s}$ for the $\nu_R$ 
at $z=0.3$.   
The direction of the flow deviates from that of the $\nu_L$, with an angle
$\langle\cos\theta\rangle=
\int\Delta_\zeta^2\theta^L_{\nu c}\theta^R_{\nu c}/k^3dk/ 
\sqrt{\int\Delta_\zeta^2(\theta^{L}_{\nu c}/k)^2d\mr{ln}k
\int\Delta_\zeta^2(\theta^{R}_{\nu c}/k)^2d\mr{ln}k}$, 
and we find $\theta \sim 20^o$.

The right-handed neutrino wake direction differs from the left-handed
wake by an angle of typically 20 deg.  This direction is
computable, and differs from patch to patch.  The signal is a small
effect on the amplitude of the total wake, which is degenerate with
measurement systematics and small variations in the left-handed
neutrino mass.  Instead, we use the difference in direction, and only
consider the right-handed wake component orthogonal to the left-handed
wake to be observable, and adjust the sensitivity estimates correspondingly.

{\it The observability.}---For forecasting purposes, 
we consider a combined LSST+Euclid data set as well as a future 21 cm lensing 
survey. The LSST will survey about three billion
galaxies and the expected error for $\kappa$ is about
$0.28/\sqrt{N}\simeq5.2\times10^{-6}$, where
$N=2.88\times10^9$ \cite{Abell:2009}. The Euclid survey 
gives a similar expected error \cite{Laureijs:2011}\cite{Amendola:2012}. 
Here, for simple estimates we will neglect the overlap between
these two surveys and assume that they provide independent data sets. 
By combining these data, we can reach a precision of
$\sigma_{\kappa_1}=5.2\times10^{-6}/\sqrt{2}=3.68\times10^{-6}$.

Ultimately, extremely high precision measurements can be achieved with
21 cm lensing surveys such as the one proposed in Ref. \cite{KiyoUeLi}.
For such a survey, the error on $\kappa$ is $(k_\mr{max}/k_\mr{min})^{-1.5}$.  
For a survey with $k_\mr{max}=1.47\times10^2\ h$/Mpc and
$k_\mr{min}=1.47\times10^{-3}\ h$/Mpc, the expected error is about
$3.16\times10^{-8}$, which is far smaller than the signals. With such
future 21 cm lensing surveys, we can measure neutrino masses with
unprecedented precision, and may even determine whether the neutrinos  
are Majorana or Dirac particles.

Based on the measurement of neutrino mass differences from the neutrino
oscillation experiments, three possible hierarchies are being considered:
(i) the normal hierarchy (NH) case, with $m_1\sim m_2 \approx 0, m_3 \approx 
0.05 \eV$; (ii) the inverted hierarchy (IH) case, with 
$0 \sim m_3 \ll m_1 \sim m_2 \approx 0.05 \eV$;
(iii) the quasidegenerate (QD) case, with
 $m_1 \approx m_2 \approx m_3 \approx 0.1\eV$. 
In the IH and NH cases, we can neglect the contribution of the 
lighter neutrinos, so the NH case is equivalent to a single neutrino 
with a mass of 0.05 eV, while the IH case is equivalent to two 0.05 eV 
neutrinos. The QD case is almost equivalent to three 0.1 eV neutrinos.

\begin{figure}[tbp]
\begin{center}
\includegraphics[width=0.48\textwidth]{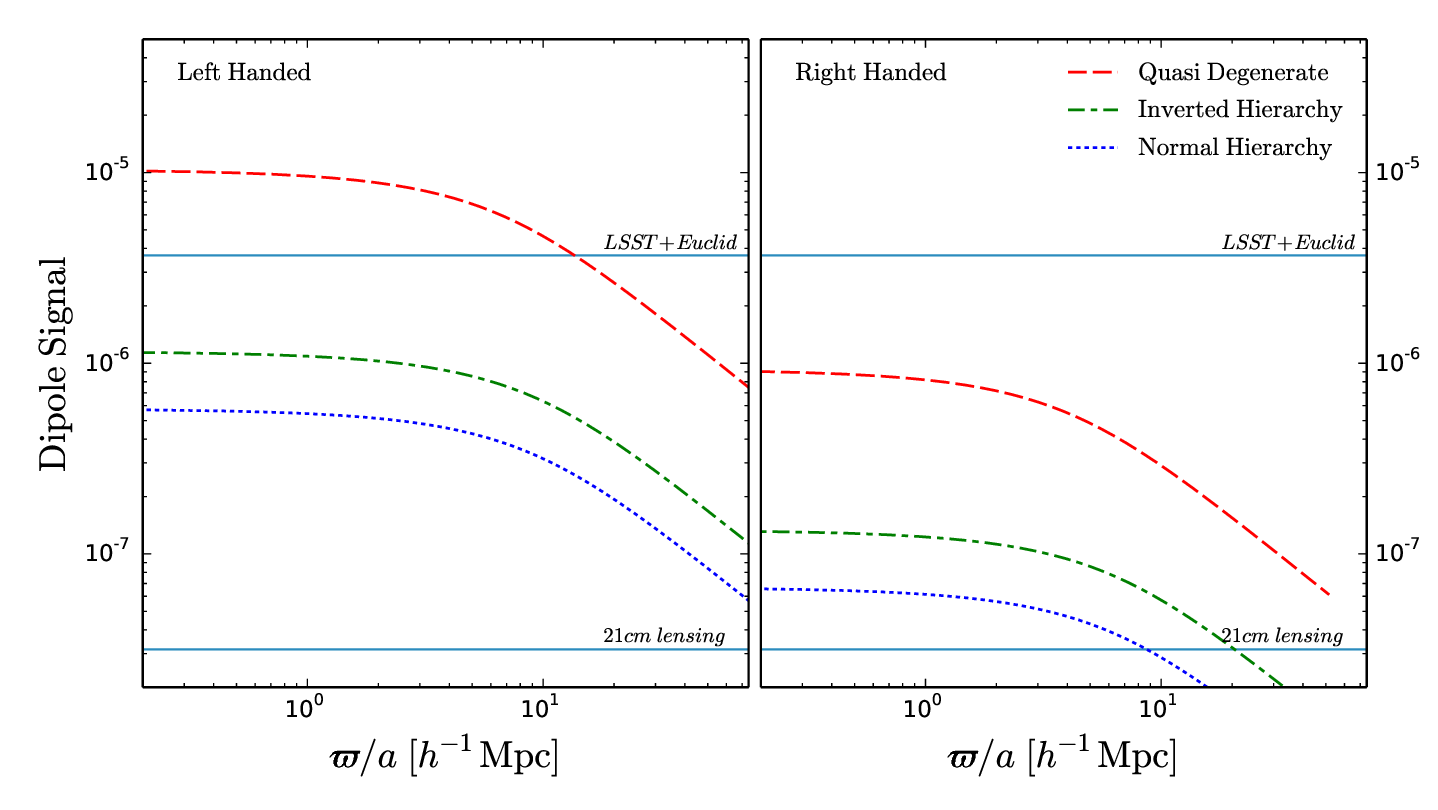}
\end{center}
\vspace{-0.7cm}
\caption{\label{fig:kappa1} Dipole signal $\bar{\kappa}_1$ 
    with consideration of neutrino hierarchy.  The noise floors
    for LSST+Euclid lensing and for 21 cm lensing are also shown. For the
    right panel, only the signal orthogonal to the left-handed
    neutrino wake is shown.
}
\end{figure}

The dipoles on the sky depend on the angle $\theta$ between the relative
velocity direction and the l.o.s.; the signal reaches its maximum when 
the relative velocity is perpendicular to the l.o.s. ($\theta=\pi/2$) and 
vanishes when $\theta=0$.
Thus, the final observed signal strength is an average over different angles.
In Fig. \ref{fig:kappa1}, we plot the expected dipole term 
$\bar{\kappa}_1=\int\kappa_1(\theta)d\theta/\int d\theta$ for the $\nu_L$ 
(left panel) and the $\nu_R$ (right panel) for these three cases.
The dipole signal for the $\nu_R$ is suppressed by $\cos 20^o$ since we
only consider the wake component geometrically orthogonal to the $\nu_L$.
We also plot the measurement errors for the
LSST+Euclid lensing survey and the future 21 cm lensing survey described
above. The LSST+Euclid data should be able to positively
detect the QD case. With the future 21 cm lensing survey, the
NH and IH cases can be distinguished, and even the $\nu_R$ 
may be detectable.

{\it Conclusion.}---We have computed the density contrast of neutrino
wakes produced by the relative motions of neutrinos and dark matter.
We have estimated the observability of these wakes via gravitational
lensing and have shown that it may be possible to observe both the mass
hierarchy as well as right-handed neutrinos.  The
wake directions
differ by a distinctive angular signature.

We acknowledge the support of the Chinese MoST 863 program under Grant 
No. 2012AA121701, the CAS Science Strategic Priority Research Program 
XDB09000000, the NSFC under Grant No. 11373030, Tsinghua University, 
CHEP at Peking University, and NSERC.

\bibliographystyle{apsrev}
\bibliography{neubib}

\end{document}